# Geo-engineering Gone Awry: A New Partial Solution of Fermi's Paradox

By

Milan M. Ćirković

Astronomical Observatory Belgrade

Volgina 7, 11160 Belgrade

SERBIA AND MONTENEGRO

arioch@eunet.yu

Richard B. Cathcart

Geographos

1608 East Broadway, Suite #107

Glendale, California 91205-1524

UNITED STATES OF AMERICA

rbcathcart@msn.com

**Abstract.** Another partial solution of Fermi's famous paradox is proposed, based on our increased understanding of geophysics, geo-engineering and climatology. It has been claimed in the recent astrobiological literature (for instance, in the recent controversial "rare Earth" theory of Ward and Brownlee), that geological activity of a terrestrial planet is an important precondition for the emergence of complex metazoan life forms. Technological civilizations arising on such planets will be, at some point of their histories or another, tempted to embark upon massive geo-engineering projects. If, for some reasons only very recently understood, large-scale geo-engineering is in fact much more dangerous than previously thought, the scenario in which at least some of the extraterrestrial civilizations in the Milky Way self-destruct in this manner gains plausibility. In addition, we speculate on possible reasons, both physical and culturological, which could make such a threat even more pertinent on an average Galactic terrestrial planet than on Earth.

## 1. INTRODUCTION

The absence of extraterrestrial intelligent beings within Earth's biosphere could be best explained by the conjecture that advanced technological aliens are non-existent in our Galaxy—as claimed by Enrico Fermi (1901-1954) at a lunch party in May 1950.[1] Assuming another Earth-like planet elsewhere in the Milky Way that is inhabited by persons similar to us in both epistemic capacities and technology, then it is necessary to imagine those particular aliens wondering why there have been, thus far, no official reports of Earthling visits! In this report we examine a variant of the possibility that intelligent extraterrestrials, natives of a terrestrial planet, have exterminated their species accidentally, ending all technological examination of their planet and its surroundings as well as macro-engineering alterations of it. To add realism's flavor to our scenario, we take as our chief example the world we know best, the Earth—and, in particular, how a self-induced geological/Macroengineering extermination event-process might result in an Earth-biosphere without human beings present. In our scenario, extant Earthlings function as stand-ins for typical aliens, according to the Copernican Principle.

During the 20th Century, a wartime global thermonuclear winter was postulated as a potential terminal event-process for *Homo s. Sapiens'* extinction.[2] Now, we postulate a future inadvertent Earth *Venuforming*—terraforming's exact opposite[3]—which will remove any possibility for humans to engage with aliens, ever. Another appropriate term is "ecocide": a peacetime activity that destroys or damages the Earth-biosphere on a massive geographical scale, ultimately making the Earth unfit for human life.[4] In this instance, we will blame a single rapidly completed macroproject, David J. Stevenson's May 2003 proposal for a "Mission to Earth's Core" (MTEC)[5], undertaken for a noble purpose by unwary and uninformed human macro-engineers. However, it is used mostly as a prototype of the sort of Macro-engineering activities intelligent beings could undertake with incomplete understanding of the potential dangers involved and/or willingness to take the risk. The potential danger belongs, as we shall elaborate below, to the class of *existential risks*,[6] to which we shall have to focus our attention in the years to come. Economists have very recently warned that "…environmental impact assessments ["impacts" are all on human society] of… macroprojects… are often restricted to check-listing procedures that stress well-established knowledge on local impacts, while ignoring interregional, global, systemic or long-term effects."[7]

Geological maps of Earth, drawn after observation of our planet's surface, form a fundamental basis for interpreting our homeland's geological history that, in turn, forms the basis for understanding operative planetary event-processes through its unique Geological Time.[8] The first planetary geologic maps were made of the Moon, a relatively tectonically inactive Solar System body. Completed geologic maps are partly fact and partly "best guess" interpretation. Earth's volume is $\sim108 \times 10^{10}$ km$^3$. Its present-day continental surface area ($\sim148 \times 10^6$ km$^2$) includes 24.7% complex exposed rock, 10.9% shale, 22% sandstone, 5.7% basalt, 18.2% granite, 8.3% carbonate and the remainder (15.1%) is ice-covered and therefore what kind of rock supports the ice is as yet unknown. Only a future-perfected seismology can inform us about the solid, liquid and gaseous substances which make up Earth's body. The interior of our planet is a legitimate target for scientific exploration, and potential future resource management, but our quite incomplete understanding of it hides very serious dangers.

## 2. MACRO-ENGINEERING AND THE CATASTROPHE SCENARIO

Sir Martin Rees, the Astronomer Royal, assigned disturbingly high probability to humanity's termination due to a catastrophe instigated by scientists or technologists in *Our Final Century: Will the Human Race Survive the Twenty-First Century?* (2003). Certainly we act together as a globalized civilization when people observe the Universe and discuss the consequences of our observations, including the most pertinent technological consequences. In the same time, this globalized civilization is threatened by a relatively recently acknowledged spectrum of *existential risks*, being (in the definition of Ref. 6) those "...where an adverse outcome would either annihilate Earth-originating intelligent life or permanently and drastically curtail its potential." Some of the very well-known existential risks are of natural origin, the prototype being collision of a large asteroid or comet with Earth. Others are of anthropogenic nature—the example of global nuclear war comes immediately to mind, but there are also many others. There are two crucial insights stemming from still scarce scholarly research on the topic of existential risks: (i) they are not subject to the standard procedures of risk analysis and management, and (ii) those existential risks of anthropogenic origin in general heavily outweight those of natural origin. These insights of Rees, Bostrom and others may be absolutely correct in his environmental impact assessment and projection, while still missing some additional anthropogenic (specifically, geologic) threats for *Homo s. Sapiens.*[9]

We hereby propose an apocalyptic scenario of greenhouse gases vented from the Earth's crust and mantle because humans undertook Stevenson's MTEC macroproject, as a prototype of these additional existential risks. Since humans are currently without means of escape from the planet (the state-of-affairs likely to persist for at least several decades and probably much longer), and their Macro-engineering experiment meets with misfortune, they will perish as a species due to the excessive heat possibly Venuforming the Earth, i.e. permanent locking in the runaway greenhouse heating. Of course, it is possible that the Earth-biosphere continues to function and that subsequently, owing to the atmosphere's increased perennial warmth[10] and the aerial fertilization of expelled $CO_2$, vegetation will be caused to flourish. Surface life forms must thereafter endure much different solar ultraviolet radiation (UVR): under heavy cloud cover the scattered UV component of sunlight—often termed "skylight"—is seldom less than 10% of that under a clear, blue sky; very heavy storm clouds can almost eliminate terrestrial UVR even in summertime.

Ironically enough, Macro-engineering projects are often suggested as schemes to successfully *mitigate* adverse effects of anthropogenic technological activities. The idea most investigated so far aims to counter the industrial emission of greenhouse gasses.[11] Although the conclusions of a recent study is that "there are many reasons why geo-engineering is not a preferred option for climate stabilization"[12], it is still very much a new professional field in which many different conclusions may be reached. It is conceivable that our civilization will be *forced* to undertake some form of geo-engineering, especially if it is confirmed that there are no natural climate forcings that are able to counteract anthropogenic destabilization of our habitat.[13] If our case is typical, according to the Copernican Principle, we are entitled to envisage a significant fraction of all extra-terrestrial civilizations undertaking Macro-engineering, with a wide spectrum of possible outcomes, some of which are bound to be catastrophic (we shall return to this topic in Secs. 5 and 6 below).

## 3. MISSION TO EARTH'S CORE

Most persons know that many geologic event-processes occur at rates similar to rates of change in everyday life.[14] However, there are event-processes that are very rapid and extra-ordinarily powerful as, for example, volcanic super-eruptions.[15] There are strong indications that such events (in particular the Toba super-eruption about 74,000 yrs ago) already almost caused humanity's extinction.[16] About 4 P.M. on 20 February 1943, in a Paricutin, State of Michoacan, Mexico maize field owned by Dionisio Pulido, a volcano commenced its lava eruption onto Earth's land surface. [On 20 December 1944, in <u>Excelsior: The Newspaper of National Life</u> (XXVIII Year-Vol. VI, No. 10,005), in Mexico City, Pulido offered his volcano for sale.] Although the Paricutin Volcano became "dormant" *circa* 1952, by 1950 a pioneering geomorphic study of gully development was possible because of Paricutin's accurately dated natural construction.[17] And, just a few years later, all the major ecological changes consequent to the surprise 1943 eruption were documented carefully.[18] Ultimately, volcanoes everywhere are powered by heat generated at Earth's core as well as above it.

Let us suppose, for the purposes of this speculation, that an anthropogenic volcano becomes active through Macro-engineering's misapplication. What might be the main effects of its appearance and constant operation?

Techno-thriller audiences watched THE CORE in cinemas everywhere during 2003; its Jules Verne-inspired storyline enthralled many science fiction fans that were favorably impressed by the film's state-of-the-art, computer-generated special effects.[19] No living scientist actually knows with categorical certainty of what material(s) Earth's core is composed and what is its current physical state. At our planet's center of mass there may be a tiny, 3 km-diameter "nuclear reactor".[20] Starting in 1943, A.J. Shneiderov speculated that Earth's core was super-dense hot plasma energized by the continuous, variable flux of cosmic particles modulated by the Sun, Moon and other Solar System planets.[21] Whatever its contents, Earth's core must influence all materials and life situated above its position inside the rotating planet. And, MTEC has as its primary goal a visit by a remote sensing device descending to the core gravitationally. It is conceived as a mission to Earth's core in which a probe is embedded in a large volume of liquid iron alloy that migrates to the core along a crack propagating under the action of gravity. In addition to at least $10^8$ kg of melted iron, a convenient crack has to be found, or even more likely, artificially made using a thermonuclear explosion of the order of several megatonnes. The completely solid-state probe would contain miniaturized instrumentation for performing geophysical measurements and for communicating its results via acoustic waves detectable by low-frequency gravitational wave detectors at the surface. (Overlooked by Stevenson, William Mansfield Adams, during 1965 offered a device to obtain a sample of mantle material and to return any obtained sample to Earth's surface.[22]) MTEC deep-Earth probe would descend at nearly 100 kph, reaching Earth's core in approximately seven days. The probe's descent can be expected to initiate remarkably unnatural geological event-processes that are exceedingly dangerous for the Earth-biosphere, potentially causing an anti-biotic crust-mantle geological upheaval.

## 4. COMMENCEMENT OF EXTINCTION EVENT-PROCESS

How could humankind's MTEC-caused extinction begin? Hans Keppler and his colleagues offer a tentative answer: mass extinctions can be caused by the liberation of carbon formerly lodged in the Earth-mantle.[23] It is already demonstrated that methane hydrates form a voluminous reservoir of carbon and the quick, punctuated release of this frozen methane—the "Clathrate Gun Hypothesis"[24]—results in global warming of the Earth-atmosphere. This might have actually happened at the end of Permian (*cca.* 251 Myr ago), causing the greatest mass extinction ever recorded in the history of terrestrial life, according to the newest study.[25] MTEC, taken as a whole, amounts to a hasty assemblage of all the necessary basic components for a "Carbon Dioxide Gas Gun Hypothesis"! If misused, Stevenson's MTEC macroproject could eliminate humans from the Earth-biosphere for-ever since the atmosphere's heating will cause reproductive sterility and bodily heat stress far beyond normal tolerance.[26] At the site of its inauguration, the "Mission to Earth's Core" probe may permit and construct a volcano-like "crust-mantle cannon" shooting unwanted—that is, biologically undesirable—greenhouse gases upwards. Interestingly enough, a well-considered speculation linking atmospheric $CO_2$ buildup to biological extinctions repeatedly appears in the literature.[27]

When Stevenson's probe encounters groundwater it will certainly cause the interaction of freshwater with magma. Earth-normal thermal equilibration between freshwater in a ratio of $10^{13}$ gram-moles of $25^0$ C freshwater to one cubic kilometer of $1200^0$ C magma then evolves $10^{12}$ gram-moles of gaseous hydrogen. Freed hydrogen gas may have a deleterious effect on Earth's atmosphere[28], before it leaves the planet.[29] Human environmentalists have consistently promoted early hydrogen gas use technology as a panacea for Earth-biosphere and international economic ills, including urban and regional air pollution as well as atmospheric warming.[30] Hydrogen's industrial usage worldwide would inject anthropogenic water into the stratosphere and, consequently, cooling of the lower stratosphere and a disturbance of the ozone chemistry; human-made hydrogen will cause Earth's atmosphere to become cloudier! At still greater depths, hydrogen may be commonly present in large quantities, all of which could interact dangerously with the Stevenson probe.[31]  During a very abbreviated period of Earth's Geological Time—namely, a recorded (historical) time period of only one week—the MTEC macroproject eventually may obviously transmogrify the Earth-mantle's constituents (pollute) and render the shear-cracked Earth-crust less naturally integrated.  (Around 10 GJ is required to vaporize a cubic meter of Earth-crust material.)  Penetration to the mantle using the MTEC device could also, perhaps, indirectly substantiate a nowadays quite nebulous geological hypothesis on a very controversial subject that was formulated by Vladimir A. Epifanov of the Siberian Research Institute of Geology, Geophysics and Mineral; at the 2002 Moscow, Russia, "Conference on Degasification of the Earth", Epifanov proposed the 30 June 1908 Podkamennaya Tunguska Explosion was not caused by a meteorite or comet's violent impact upon the ground surface but, rather, by a fluid carbohydrate jet which shot upwards under high pressure from a large subterranean reservoir!  Epifanov assumed the massive, and radiative tree felling in that almost uninhabited Siberian region was chiefly instigated by atmospheric pressure waves caused by an aerial gas explosion (fireball).

## 5. IMPLICATIONS FOR ASTROBIOLOGY

Fermi's Paradox is the contradiction between the apparent absence of recognizable aliens, and the common expectation that humans ought to observe and/or hold in their hands some perceivable evidence of alien existence. It has been realized long ago that the straightforward explanation for the "paradox" is that the intelligent communities in the Galaxy are too short-lived for communicating or colonizing over the typical interstellar distances. The obvious cause of such short life in the communication phase is succumbing to an existential risk (appropriately generalized to apply to any intelligent life). In particular, those risks incurred by irresponsible use of science and technology are what we can call "suicidal" scenarios. In fact, Stevenson's "Mission to Earth's Core" could be considered a new form of "suicidal" type solution—additional to thermo-nuclear Armageddon or nanotechnological "gray goo" catastrophe—for Fermi's paradox. In the terrestrial context the scenario is quite simple: our globalized civilization overdevelops Macro-engineering earlier than, say, colonization of other planets, nanotechnology, or advanced forms of biotechnology[32], and as a direct follow-on destroys our species by a misguided intra-Earth exploration.

The idea that geological event-processes may influence the tally of extra-terrestrial civilizations receives further support from some recent astrobiological results.[33] Notably, Peter Ward and Donald Brownlee argue in their influential monograph *Rare Earth* that geological activity in general, and plate tectonics in particular are *essential* for the development of complex metazoan life on *any* Galactic planet.[34] If that claim is correct, then natural preconditions for the catastrophic scenario outlined above are *a fortiori* present for any technological species; the total frequency of such locations (the most controversial claim of the Ward and Brownlee theory) is unimportant from our present point of view.

One step further would be to conclude that geological activity of terrestrial planets has, in fact, been stronger in the Galactic past. This is due to the increased chemical abundances of radioactive elements, notably U, Th, and isotope $^{40}$K, main generators—other than, of course, incident solar energy—of planetary geological activity and its interaction with the atmosphere. It has been shown recently that most terrestrial planets in the Milky Way Galaxy are significantly older than Earth; in fact, the average age of a terrestrial planet is $6.4 \pm 0.9$ Gyr, according to the calculations of Lineweaver.[35] Thus, the oldest civilizations should have arisen in the epoch in which the cosmic abundance of radioactive supernovae-created elements was higher than today, and thus the terrestrial planets were, on average, more geologically active. It is precisely the oldest civilizations that present the biggest problem from the viewpoint of Fermi's paradox! The "Great Silence" would be explained more easily if we assume that the "mortality rate" of technological societies was higher in the past, anti-correlated with the amount of average geological activity. Part of this story perhaps resides with the natural geological causes, notably super-volcanism mentioned above; but another part may lay with technogenic geological catastrophes.[36]

It is important to emphasize that this proposition is suggested only as a *partial* solution to Fermi's paradox; it complements other "self-destruction" hypotheses (many of which are listed in the Ref. 1), such as those having extra-terrestrial civilizations destroying themselves through thermo-nuclear wars (and subsequent global "Nuclear Winters"), or nanotechnological "gray goo". It would be naive to claim that any *single* cause is the solution for the absence of extra-terrestrials; rather, it would be some combination of causes, including both "gradual" (like various transcendence/devolution scenarios, or even the "Zoo"[37] and "Interdict"[38] hypotheses) and "catastrophic" scenarios. Obviously, it is of foremost importance to investigate the latter, since it is those that will reveal to us aspects of both hypothetical aliens and, quite probably, humankind's future prospects.

## 6. CULTUROLOGICAL AND CLIMATOLOGICAL FACETS

One cannot avoid speculation upon possible reasons leading or compelling technological civilizations to undertake potentially hazardous geo-engineering. These reasons belong to the realm of sociology of science and culturology, and thus are bound to be speculative excursions focused on the human mind or human-like alien mind. The argument of this Section should, therefore, be taken with caution.

Revolutionary impact of astronomical observations on the history of science and technology here on Earth has been noticed and investigated in various contexts. The best-studied case is, undoubtedly, the Copernican Revolution[39] which led directly to modern science and technology, but there is a plethora of others. It has not been said in vain that "…it was really the Moon that provided the principal ideas as well as the crucial tests for our understanding of the Universe".[40] On an even broader cultural basis, one may cogently argue for characterization of human civilization(s) as essentially "upward-oriented". In most of human cultures at all epochs, deities inhabit mountain peaks, sky, specific celestial objects or abstract "heavenly" realm. Most holy places of humanity are either topographically elevated in order to be closer to the sky (Lhassa, Mt. Olympus, Sinai, Mt. Athos, Teotihuacan) or associated with the upward motions/ascensions (The Dome on the Rock in Jerusalem, European cathedrals, Buddhist pagodas, Islamic minarets). The nascent science of Astro-archaeology traces the influence of celestial bodies on human life and culture for at least 6,000 years.

The necessary precondition for all this is, obviously, the fact that Earth's atmosphere is largely transparent at tropical and continental mid-latitudes (or at least it was before the advent of the industrial age and both aerosol and urban nighttime light pollutions.) A United Nations Organization scientific inquiry, Project Asian Brown Cloud, is underway to examine the spreading, fuliginous atmospheric shroud over India and China. What if humans, with perfect vision, could never see the Moon or the Universe's ~6,000 stars viewable from Earth?  Let us suppose Earth's sky was a thick deck of clouds through which only hazy sunlight could penetrate, as a distinguished mathematician E. Brian Davies[41] has with a speculative historical essay "The Role of Astronomy in the History of Science".[42]  Stephen Webb offered "Solution 29: Cloudy Skies Are Common" in cited Ref. 1 as a solution why some or most of extra-terrestrial civilizations do not have the means of communicating with us. We find his asserted idea that such dense cloud cover *entirely* prevents the emergence of science and technology rather extreme and unsupported.  Prof. Davies, on the other hand, defends the contrary scenario: namely that science and technology will develop, albeit slower and somewhat differently, even in the absence of astronomical observations and the challenge they presented for wise men during the millennia. However, a possible catch lies in the nature of that difference. Civilizations arising on planets with dense cloud covers (similar to those shrouding the surfaces of Venus and Titan) will presumably be more "downward-oriented". They will tend to develop geosciences at the expense of astronomy and related sciences. Financial and material resources of such a civilization will be allocated to travel downwards rather than upwards. Perhaps equivalents of NASA, ESA, and other space agencies will, in that case, spring into existence with purpose of exploting the interior of the globe instead of its cosmic surroundings. This would have a profound impact on the possibility and likelihood of their developing potentially dangerous geo-engineering. In addition, a lack of clear example of the greenhouse effect (the historical role Venus plays for our civilization) will probably decrease awareness of such threats in both scientific and laymen's circles.

(This is not to say that the influences from below were unimportant in the course of human history. It is worth noting that a succession of Greek maidens, induced into a revelatory trance by various gases emanating from the earth below the Delphic oracle temple site, scouted the course of Greece's history![43] Today, there is an increasing world-public interest in subterranean living.[44] On the other hand, this seems to indirectly corroborate the idea of Davies that advanced cultures which include science can, in principle, form under "downward oriented" influences.[45])

We speculate that a typical global human civilization of a look-alike Earth will be prompted to penetrate their Earth's subsurface vigorously since that is the only physical realm open to easy preliminary exploration; furthermore, we speculate that this collectively intelligent group would enthusiastically endorse and practice Macro-engineering.[46] Ultimately, invention of aerospace plane technology would allow people to appreciate that "The heavens are telling the glory of God" (Psalm 19:1), since they could then move into the essentially limitless continuum extending from Earth's surface outwards through the atmosphere to the farthest parts of the observable Universe, especially embracing attainable parts of the Solar System.[47] But, since even present-day humans have no such flying aerospace vehicles, it seems reasonable to ignore the implications of their absence or existence. There is a formulaic classical phrase often appearing in epic ancient poetry which labels the borderless ocean "misty" or "airy" suggesting, obviously, a "cloak of fog", which is a cloud that touches the ground's surface. Some macro-engineers take this artistic Greek phrasing to be a comment on Earth's Hydrologic Cycle, water vapor circulating, in part, in Earth's atmosphere. But, to our purpose, the ancients also used clouds as designated physical boundaries for the known world (the Greeks' *oecumene*). Our report deals with a horizontal planet-engulfing cloud, while our ancestors of two millennia past had to imagine a peripheral (Mediterranean Sea Basin) vertical cloud preventing—or at least usefully demarcating—their ship-borne explorations of the eastern Atlantic Ocean![48]

Our Earth's current natural cloud coverage ranges from ~65% to 68%, plus or minus 4.8%. Galactic cosmic particles—especially the solar cosmic particles emanating from the Sun's chromosphere—collide with other particles in the atmosphere and thereby cause some clouds to form. In other words, the Sun modulates Earth's sky![49] Important *additional modulation* can come, it has been recently claimed, from the Galactic cosmic-ray sources, concentrated in the Milky Way's spiral arms and central (bulge) regions.[50] Thus, it is possible that there are entire vast regions of the Galaxy (sub-regions of the Galactic Habitable Zone) in which predominantly cloudy terrestrial planets prevail, and most arising civilizations are of this, "downward-oriented" type.

How can this scenario be empirically tested? As with most "local" explanations of Fermi's paradox, it is a rather difficult task to perform, but there are a few clues. The detection of terrestrial planets around other stars remains probably the most important observational task in astronomy of the next decade; several ambitious projects, including the DARWIN[51] and GAIA[52] space-based observatories, are designed for this exact purpose. Since astronomers have already investigated the atmosphere of a gigantic Jovian extra-solar planet, it is not unreasonable to expect that future observations will offer humans a glimpse of extra-Solar System terrestrial planet geophysical properties. This will enable us to investigate how commonplace cloudy planet skies are throughout the Galaxy. On the other hand, we certainly would not wish to empirically test the safety of Stevenson's MTEC device and all similar contraptions; however, non-invasive geophysical studies will certainly shed more light on the issues of carbon release, and effects of perturbations on the delicate known mantle-crust-atmosphere feedbacks.[53] Finally, it is to be hoped that future studies in sociology of science and technology will reveal to humans precisely how typical our relatively poor understanding of geosciences is in comparison to a set of model cultures. In this manner, the "geo-catastrophe" scenario is more amenable to empirical verification than most other similar scenarios for resolution of Fermi's paradox.

## 7. CONCLUSIONS

We present a novel "catastrophic" solution to the problem of absence of advanced extra-terrestrial civilizations or their manifestations ("Fermi's paradox"). In addition to already discussed variations on the topic of technological traps awaiting intelligent communities—like the nuclear or nanotechnological catastrophes—we point that advanced geo-engineering can, for accidental or intentional reasons, bring the biological downfall of such remote industrialized communities. Only very recently we have reached the stage in which such a macroproject, Stevenson's "Mission to Earth's Core" (MTEC) device, can be technologically achieved. Besides suggesting increased public attention to this producible new existential risk, we suggest that this fate has befallen and can potentially befall at least some other Galactic civilizations. We indicate some, still very speculative, reasons why this climatic calamity can actually be more serious on inhabited planets other than Earth. When this existential threat is added to other better-known threats, it is just possible that the total risk function facing civilizations is high enough to explain the total absence of their manifestations.